\documentclass[num-refs]{wiley-article}

\usepackage{lineno}
\usepackage{siunitx}
\usepackage{physics,amsmath}
\usepackage[caption=false]{subfig}
\usepackage{placeins}
\usepackage{xspace}
\usepackage{multirow}
\graphicspath{{Figures/pdf/},{Figures/png/}}
\DeclareGraphicsExtensions{.pdf,.png}
\def\MnCl{\ensuremath{\mbox{MnCl}_2}\xspace}

\def\TR{\ensuremath{T_R}\xspace}
\def\TI{\ensuremath{T_I}\xspace}
\def\TE{\ensuremath{T_E}\xspace}

\def\T#1{\ensuremath{T_{#1}}\xspace}
\def\R#1{\ensuremath{R_{#1}}\xspace}
\def\etal{\textit{et al.\ }}

\papertype{preprint - Submitted to Magnetic Resonance in Medicine}

\title{Comparison of \R1 Mapping Protocols: What are we measuring?}

\author[1,2,\authfn{1},\authfn{2}]{Christopher Jenkins, PhD}
\author[1,2,\authfn{1}]{Ioannis Papadopoulos, PhD}
\author[1,2,\authfn{1}]{S.~M.\ Shermer, PhD}

\corraddress{S.~M.\ Shermer, PhD}
\corremail{lw1660@gmail.com}

\presentadd[\authfn{2}]{Cardiff University Brain Research Imaging Centre (CUBRIC), Maindy Rd, Cardiff, CF24 4HQ, United Kingdom}
\affil[1]{College of Science (Physics), Swansea University, Swansea, SA2 8PP, United Kingdom}
\affil[2]{Centre for Nanohealth \& Clinical Imaging Unit, Institute for Life Science, Swansea University, Swansea, SA2 8PP, United Kingdom}

\contrib[\authfn{1}]{Equally contributing authors.}

\fundinginfo{Royal Society Leverhulme Trust (SMS); ABMU Health Board, College of Science (CJ, IP)}

\runningauthor{Jenkins et al.}

\begin{document} 

\maketitle
\begin{abstract}
\textbf{Purpose}\\
Recent work highlights the breadth of reported spin-lattice relaxation rates (\R1) for individual tissues. One potential source of variation is the protocol used to determine \R1.  The methodological dependence of \R1 and relaxivity $r_1$ are investigated.
	
\noindent\textbf{Methods}\\
\R1 is quantified in gel phantoms with varying concentration of \MnCl, and a small cohort of three healthy volunteers using different acquisition methods.  Siemens inversion recovery (IR) and saturation recovery (SR) protocols are applied to phantoms and volunteers.  Variable flip angle (VFA) protocols are additionally applied to phantoms. \R1 is quantified using single voxel fits, and distributions examined for regions in the thalamus, and cerebellum as well as grey and white matter. Phantoms exclude boundary fits and relaxivity is quantified across the full concentration range.  Normality of \R1 distributions is assessed by Kolmogorov-Smirnov score, and inter-sequence agreement by two-sample t-test.
	
\noindent\textbf{Results}\\
Phantom relaxivity is found to be \SI{7.16}{Hz/mM}, \SI{9.22}{Hz/mM} and \SIrange{10.65}{11.91}{Hz/mM} for IR, SR and VFA methods, respectively.  \textit{In vivo} \R1 exhibit low intra-participant variation for IR. SR $R_1$ are lower than IR values with inter- and intra-participant variation on the same order.  Brain regions and phantoms mapped with different protocols varied significantly with t-test $p$-values between $0$ and \SI{5e-10}{}. 
	
\noindent\textbf{Conclusion} \\
Results suggest a significant protocol dependence of \R1, and corresponding relaxivity, suggesting inter-method comparisons should be attempted tentatively, if at all.
\end{abstract}


\section*{Introduction}\label{S:intro}

Tissue relaxation rates are of fundamental importance in magnetic resonance imaging. Differences in longitudinal and transverse relaxation rates, \R1 and \R2, for different tissues are the primary source of image contrast in conventional \T1 and \T2 weighted imaging (T1w, T2w)~\cite{Elster1988}.  Differences in longitudinal relaxation rates are also exploited to suppress unwanted signals such as selective nulling of the signal from water or fat using inversion recovery sequences~\cite{Bydder1985}.  Although these techniques use relaxation weighting, the optimal choice of sequence parameters requires accurate knowledge of tissue relaxation rates~\cite{Yokoo2010}.  Absolute quantification of relaxation rates can also be advantageous when compared to relaxation-weighted techniques to minimise or eliminate potentially confounding effects of other variables such as variations in proton density or hardware factors such as field inhomogeneity or coil sensitivity~\cite{Deoni2007, Wright2008, Ma2013, Liberman2013, Cloos2016}.

Relaxation times are also biomarkers in their own right.  Quantification of \R1, which is the main focus on this paper, has many applications including cardiac \R1 mapping~\cite{Karamitsos2013,Messroghli2007} for myocardial pathology~\cite{Ugander2012} and \R1 mapping for liver fibrosis~\cite{Hoad2015} and cirrhosis~\cite{Heye2012}.   Other important applications for \R1 mapping include dynamic contrast enhanced (DCE) MRI, where modelling of contrast agent uptake in tissue, e.g., based on the Tofts model~\cite{Tofts1999} used in the prostate cancer imaging~\cite{Fennessy2012}, requires \R1 mapping.  A related application is the assessment of the safety of Gadolinium-based contrast agents (GBCA), which has recently been called into question due to evidence of gadolinium (Gd) deposition in tissue such as skin, bone and brain~\cite{Kanda2015,McDonald2015}.  Among many other applications, accurate \R1 mapping could also be a useful tool to investigate Gd deposition non-invasively by mapping tissue \R1 values over time.

However, there are large discrepancies in the \R1 values reported in different studies.  A recent review by Bojorquez \etal shows that \R1 appears to vary not only by tissue type, as expected, but can also vary by up to a factor of two for grey and white matter~\cite{Bojorquez2017}.  Some of the differences in measured \R1 rates may be the result of physiological differences in the volunteer populations studied, and small population sizes may exacerbate this problem.  Furthermore, while \R1 rates should in theory be intrinsic tissue parameters, there are many potential external factors that can affect the results, from hardware considerations such as the field strength of the scanner~\cite{Bottomley1984, Ethofer2003}, to the choice of mapping sequences and analysis procedures~\cite{Stikov2015}.  With regard to the former, in multi-site studies by Deoni \etal \cite{Deoni2008} and Bane \etal \cite{Bane2018}, the intra-site variation of \R1 and \R2 values was found to be on the same order as inter-site variation.

The focus of this paper is to examine the latter source of variation, the effect of the \R1 mapping protocols.  While a wide variety of \R1 mapping protocols have been used in published studies, the majority of them fall in one of three categories, which will be compared in this paper: inversion recovery (IR), saturation recovery (SR) and fast acquisition variable flip angle methods (VFA). \R1 quantification using IR mapping is often considered the most accurate method for \R1 quantification, followed by saturation recovery methods~\cite{Bojorquez2017}. However, both of these methods require repeated acquisitions while varying \TI or \TR respectively, rendering them time consuming and impractical for most clinical applications.  This has lead to the development of fast acquisition methods based on varying the flip angle.  One of the first, the Look-Locker inversion recovery sequence~\cite{Look1970}, decreases the acquisition time of a standard inversion recovery sequence by applying several low flip angle pulses to enable multiple signal acquisitions within a single repetition time.  Another commonly employed \T1 weighted imaging method, which was not considered in this study, is MPRAGE~\cite{Brant-Zawadzki1992}, or its more recent incarnation MP2RAGE~\cite{Marques2010}, which offers improved insensitivity to $B_1$, among other advantages.  MPRAGE uses inversion preparation pulses, followed by rapid gradient echo sampling to acquire \T1-weighted images quickly.  More recent VFA methods~\cite{Homer1985, Jiang2015} reduce acquisition times even further with ultra-fast spoiled gradient echo measurements.

The goal of this paper is to compare the three main types of \R1 mapping sequences to assess the influence of acquisition protocol upon R1 quantification. Protocols are tested for a set of \R1 contrast phantoms and several volunteers under controlled conditions designed to minimise effect of other factors, such as hardware issues or physiological differences.

\section*{Methods}\label{S:methods}

\subsection*{Test Objects and Subjects} \label{SS:Objects}

Three types of test objects/subjects are considered in this study: contrast agent solutions, tissue-mimicking gel phantoms and brain tissue.  To prepare the former, a concentrated solution of manganese chloride was made by dissolving \SI{0.1}{\g} manganese chloride (\MnCl) tetrahydrate (98\%) (CAS-13446-34-9)\footnote{Atom scientific\textsuperscript{\textregistered}} in \SI{1}{L} of deionised water.  Mass was measured using an electronic balance with a nominal precision of \SI{1}{\milli\gram}.  Allowing for other sources of error, the uncertainty in \MnCl{} measurements is estimated to be \SI{5}{\milli\gram}.  The concentrate was then further diluted, to varying degrees, to create a series of seven \SI{100}{\milli L} solutions with \MnCl concentrations of 0, 10, 20, 30, 40, 50, and 60 \si{\milli\g\per L}, where \SI{10}{\milli\g\per L} corresponds to a \MnCl concentration of approximately \SI{0.05}{mM}. Once the scan protocols had been applied to the solutions, \SI{1}{g} of agar~\footnote{Food grade agar, Special ingredients\textsuperscript{\textregistered}} was added to each solution, to create 1\% agar gels. \SI{0.1}{\g} of diazolidinyl urea (CAS-78491-02-8)\footnote{Sigma Aldrich\textsuperscript{\textregistered}} was also added as a preservative at this stage.  The solutions were heated to approximately \SI{90}{\celsius} and stirred until the agar was fully dissolved and then allowed to cool and set overnight before being re-scanned using identical protocols.  To obtain \textit{in vivo} data, three healthy volunteers, two males and one female aged 24-32, were recruited to undergo brain scans with different \R1 mapping protocols.  All participants provided informed consent before their involvement in the study. One volunteer returned for a second session the following day, where the protocol was repeated.
	
\subsection*{MR Protocols}\label{SS:Protocols}

\begin{figure}\centering
\subfloat[Inversion recovery with adiabatic inversion.]  {\includegraphics[width=0.45\linewidth]{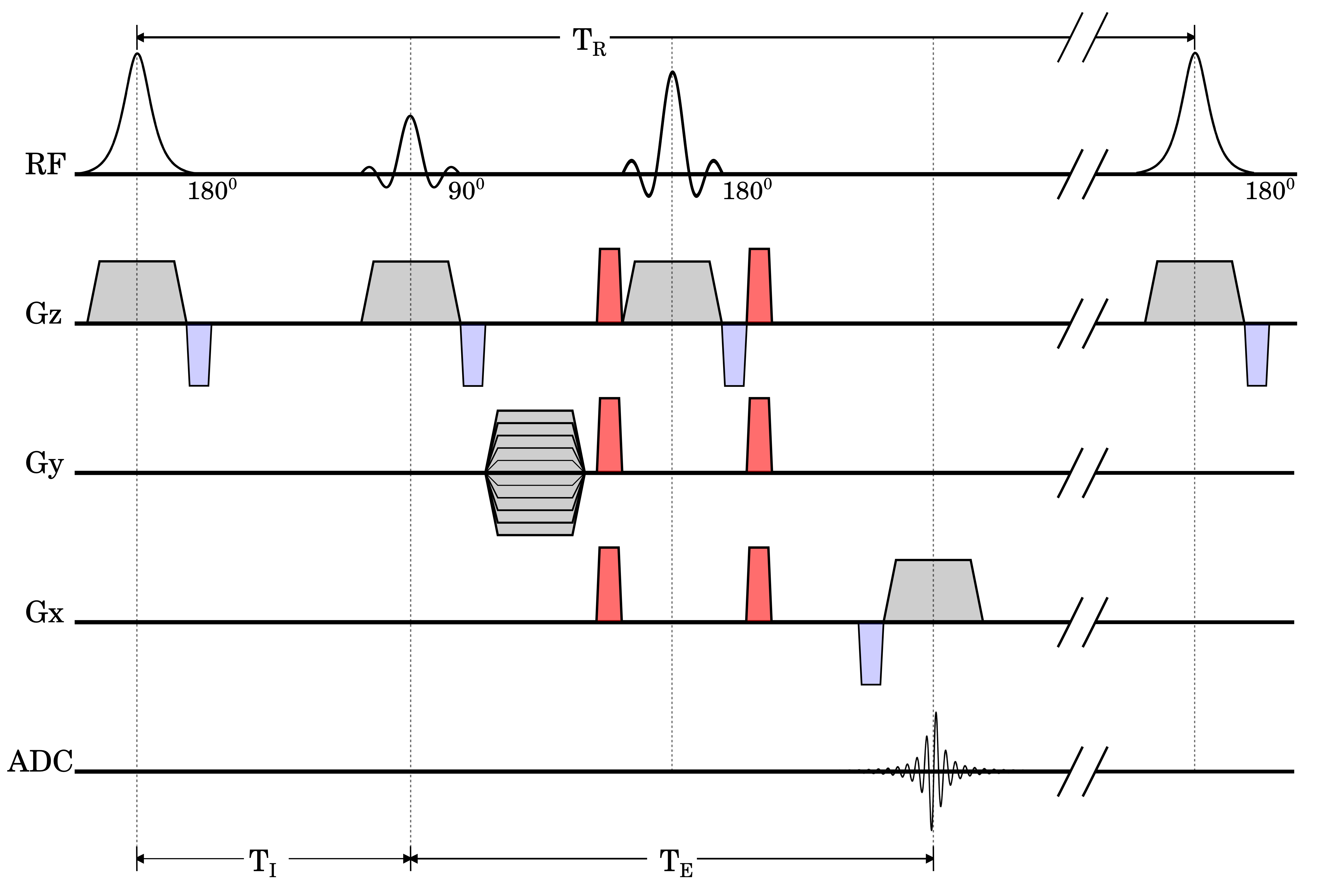}}
\subfloat[Saturation recovery.]{\includegraphics[width=0.45\linewidth]{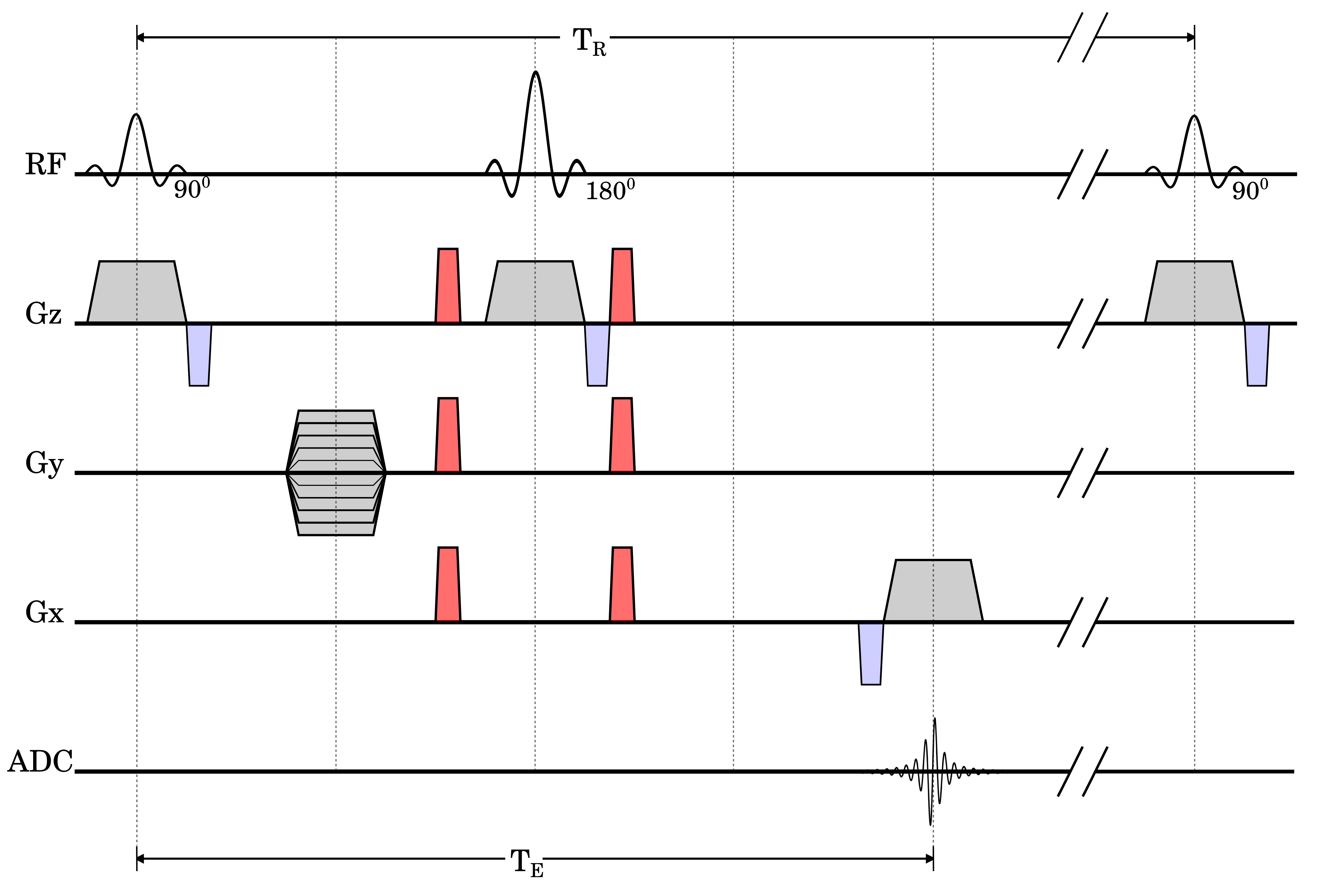}}
\caption{Pulse sequence diagrams for inversion and saturation recovery sequences.} \label{fig:Sequnces}
\end{figure}

All scans were conducted on a MAGNETOM Skyra 3 T (Siemens Healthcare GmbH, Erlangen, Germany) at Swansea university.  The scanner is situated in an air-conditioned environment was kept at a temperature of $20\pm 1.5$\si{\celsius}.  Manual shimming and transmitter calibration were performed prior to all data acquisition scans to minimise effects of $B_0$ and $B_1$ inhomogeneity. 

The seven phantoms were placed in a hexagonal arrangement around the isocentre of the magnet above the four channel spine coil element SP2.  For the IR and SR protocols a single \SI{2}{mm} thick coronal slice through the centre of the phantoms was selected and all images were acquired with a $\SI{256}{mm} \times \SI{256}{mm}$ field of view (FOV), matrix size $128 \times 128$, pixel bandwidth \SI{130}{Hz/px}, 100\% phase resolution, and one average.  The SR protocol consisted of acquiring a series of images with a vendor-supplied 2D spin echo (SE) sequence, comprised of a $90^\circ$ excitation pulse followed by a $180^\circ$ refocusing pulse with fixed \TE of \SI{12}{\milli\s}, and \TR = \SIlist{75; 125; 250; 500; 750; 1000; 1500; 2000; 3000; 4000; 5000}{\milli\s}.  For the IR protocol a vendor-supplied 2D SE sequence with an additional inversion pulse prior to the excitation pulse was used, with a fixed \TR of \SI{2500}{\milli\s}, \TE of \SI{15}{\milli\s}, and \TI values of \SIlist{30; 70; 100; 150; 300; 700; 1000; 1500; 2000}{\milli\s}.  The VFA protocol consisted of applying a vendor-supplied 3D spoiled gradient echo (GE) sequence with \TR of \SI{4.09}{ms} and \TE of \SI{1.39}{ms} for a range of flip angles between $2^\circ$ and $90^\circ$.  Twenty \SI{2}{mm} thick coronal slices were acquired and a slice through the centre, matching the slice chosen for the SR/IR protocols, was selected for the analysis.  The FOV was $\SI{256}{mm}\times \SI{256}{mm}$, matching the IR/SR protocol, while for the matrix size and readout bandwidth the default values from a clinical protocol of $192\times 192$ and \SI{390}{Hz/px}, respectively, were used.  One series was also acquired with a longer \TR of \SI{14}{ms} for comparison.  For each flip angle $8$ averages were acquired.

The volunteers were scanned in a 32 channel phased-array head coil.  To ensure coverage of the entire head, 21 slices with thickness and spacing of \SI{5}{\milli\m} were acquired.  The imaging plane was offset from transverse to coronal by \SI{25}{\degree}.  The FOV was $256\times 192$ \si{\milli\meter} and the matrix size $128\times 96$.  $96$ phase encoding steps and a per pixel bandwidth of \SI{130}{\hertz\per px} were used.  The SR protocol used a vendor-supplied SE sequence with fixed \TE of \SI{15}{\milli\second} and \TR of 286, 350, 550, 885, 1000, 2000, 4000~\si{\milli\second}.  The IR protocol used a vendor-supplied SE sequence with an additional inversion pulse with \TR of \SI{2500}{ms}, \TE of \SI{15}{ms} and inversion times of 100, 400, 700, 1000, 1300, and 1600~\si{\milli\second}.

\subsection*{$R_1$ estimation}

\begin{figure}[!htb]\centering
\includegraphics[width=1\linewidth]{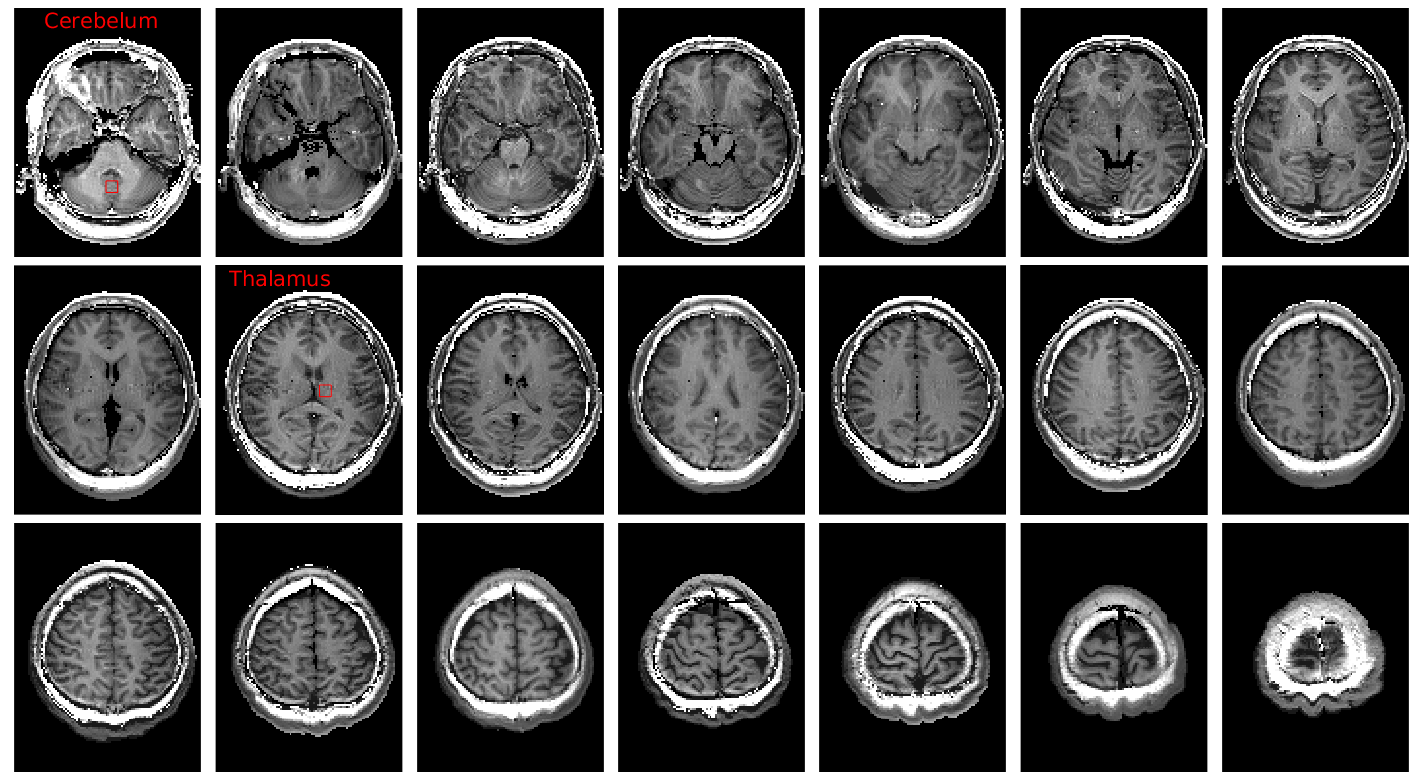}
\caption{Delineated regions of interest for one of the patient cohort. One slice is selected to delineate the cerebellum, and a second for the thalamus.}
\label{fig:brain-ROI}
\end{figure}

In-house developed Matlab code was used to analyse the scan data.  For the phantom data, regions of interest (ROIs) were delineated by thresholding the signal from one reference image, identifying the connected components and fitting circular ROIs for each component.  Approximately 10\% of the phantom's extent was excluded to limit signal variation at the interface with the phantom's container, and both single voxel and mean-signal-over-ROI fits were performed.  For the brains, square ROIs corresponding to the cerebellum and thalamus were selected manually, as illustrated in Fig.~\ref{fig:brain-ROI}.  Grey and white matter regions were identified for a representative slice in the centre of the brain using a combination of thresholding and connected component analysis to delineate grey and white matter regions for separate analysis.  For both ROI methods, a single voxel analysis was performed.

SR and IR curve fitting was carried out with a trust-region-reflective least-squares minimisation algorithm, using the Matlab curve fitting toolbox with custom fit functions. The fit function for the SR protocol was based on the formula for the transverse magnetisation as a function of \TR and \TE, derived from the Bloch equation:
\begin{equation} 
   M_{xy} = M_0 \left( 1 - e^{-\TR\R1} \right) e^{-\TE \R2}
   \label{eq:SR_Full}
\end{equation}
where $M_0$ is the equilibrium magnetisation.  As \TE is fixed in the protocol $e^{-\TE/\T2}$ was absorbed into a fit coefficient $S_0=M_0 e^{-\TE \R2}$.  To account for flip angle imperfections resulting in less than 100\% transverse magnetisation, an additional parameter $r$ was introduced, giving the final fitting function:
\begin{equation} 
   M^{\rm SR}_{xy}(\TR) = S_0 \left( 1 - r e^{-\TR\R1} \right),
   \label{eq:SR_fit}
\end{equation}
from which the parameters \R1, $S_0$ and $r$ were determined for a sequence of images with known \TR, and $M_{xy}^{\rm SR}$ determined by the signal intensity of the DICOM image.  For the IR protocol, assuming flip angles of \SI{90}{\degree} and \SI{180}{\degree} for the excitation and inversion pulses, respectively, the fit function used was~\cite{Bojorquez2017}:
\begin{equation} 
	M_{xy}^{\rm IR}(\TI) = S_0 \left( 1 - r e^{-\TI\R1} + e^{-\TR\R1} \right) 	         
    \label{eq:IR_fit}
\end{equation}
with $S_0 = M_0 \exp(-\TE\R2)$ and $r$ being a parameter to account for imperfect inversion pulses.  $M_{xy}^{\rm IR}$ is extracted from the magnitude of the DICOM signal intensity, and \TI, \TR{} are known. $S_0$, $r$, and $R_1$ are determined as fit coefficients. The quality of each fit is assessed using the coefficient of determination, $R^2$, defined as the ratio of the sum of squares of the regression to the total sum of squares.  $R^2$ can range between zero and one, with higher values indicating better fits. 

For the VFA methods, we rewrite the equation for the transverse magnetisation for a spoiled gradient echo with \TR and flip angle $\alpha$:
\begin{equation}
  M_{xy}(\alpha) = S_0 \frac{\sin\alpha}{1-E \cos\alpha},
\end{equation}
with $E = \exp(-\TR \R1)$ as a system of linear equations in $S_0$ and $E$ and define an error vector:
\begin{equation}
  \vec{e} = \begin{pmatrix} \sin\alpha_1 & M_{xy}(\alpha_1) \cos\alpha_1 \\
  \vdots & \vdots \\ \sin\alpha_n & M_{xy} (\alpha_n) \cos\alpha_n 
  \end{pmatrix} \begin{pmatrix} S_0 \\ E \end{pmatrix}
  -\begin{pmatrix} M_{xy}(\alpha_1) \\ \vdots \\ M_{xy}(\alpha_n) \end{pmatrix}.
\end{equation}
If there are exactly $m=2$ different flip angles then there is generally a unique solution with $\vec{e}=\vec{0}$.  For $m>2$ the system is over-determined and we minimise the least-squares error or $L_2$ norm of the error vector $\vec{e}$ to determine $S_0$ and $E$ from which we can infer $\R1 = -\ln(E)/\TR$.  In this case the quality of the fit is determined by the $L_2$ norm of $\vec{e}$.  

\subsection*{Statistical Analysis}

For statistical purposes, single voxel fits were performed for each voxel inside the selected ROIs for both phantom and brains.  This results in a distribution of \R1 values for each ROI.  Assuming the values are normally distributed we can calculate a mean and standard deviation of \R1 each ROI.  We test the distribution for normality by calculating the Kolmogorov-Smirnov~\cite{Massey1951} score.  The overlap between the histograms for different ROIs yields information about the distinguishability of the \R1 values for different ROIs.  To establish whether there is a statistically significant difference between the means in the \R1 distributions obtained for the same phantom or brain ROI with different \R1 mapping methods, a two-sample t-test~\cite{Johnston1970} is performed and $p$-value and confidence interval for the estimated difference in the means for the \R1 distributions is determined.

For the \R1 contrast phantoms, both liquids and gels, \R1 should increase linearly with the concentration of the contrast agent.  Therefore, we also plot the mean and standard deviation of \R1 as a function of the concentration of \MnCl and perform a linear fit to determine the slope of the line, which corresponds to the $r_1$ relaxivity of the contrast agent, enabling us to compare not only the \R1 values but also the relaxivity of contrast agent obtained for different \R1 mapping methods.  

\section*{Results} \label{S:results}

\subsection*{Phantoms}

The \R1 values for the liquid and gel phantoms obtained with different methods are tabulated in Tables~\ref{table:R1-IR-SR-compare} and \ref{table:R1-VFA-compare}.  The mean and standard deviation were calculated from the distributions of the single voxel fits for each phantom.  The tables clearly show significant variation in the \R1 values obtained for the \emph{same} phantom using different methods.  To further elucidate the magnitude and significance of the differences in the \R1 means, we performed pairwise t-tests on the \R1 distributions obtained with different methods for a given phantom. The results tabulated in Table~\ref{table:CI-R1-ttest}.

Comparing the SR and IR protocols, the differences in the means of the distributions were larger for phantoms with lower concentrations of \MnCl.  For the VFA method, the differences in the \R1 means for different VFA protocols were smaller, but they differed significantly from the estimates obtained with the SR and IR protocol.  For phantoms with low concentrations of \MnCl, the \R1 values obtained for the VFA methods, especially the common two-point method with flip angles $2^\circ$ and $15^\circ$, were much closer to SR values than IR values, while the situation was reversed for higher concentrations of the contrast agent. 

\begin{table}[!htb]\centering
\scalebox{0.75}{
\begin{tabular}{|c|cc|cc|cc|cc|} \hline
& \multicolumn{2}{c|}{IR - solution} & \multicolumn{2}{c|}{SR - solution} & \multicolumn{2}{c|}{IR - gel} & \multicolumn{2}{c|}{SR - gel}\\\hline
\textbf{\MnCl{} Conc} & \textbf{$\overline{\R1}$} & \textbf{$\sigma$} & \textbf{$\overline{\R1}$} & \textbf{$\sigma$} & \textbf{$\overline{\R1}$} & \textbf{$\sigma$} & \textbf{$\overline{\R1}$} & \textbf{$\sigma$}\\
(\si{mM}) & (\si{\per\second}) & (\si{\per\second}) & (\si{\per\second}) & (\si{\per\second})&(\si{\per\second})&(\si{\per\second})&(\si{\per\second})&(\si{\per\second})\\	
\hline
		0.00 & 1.53 & 0.033 & 0.27 & 0.017 & 1.58 & 0.020 & 0.96 & 0.027 \\
		0.05 & 1.68 & 0.028 & 0.71 & 0.030 & 1.68 & 0.12 & 1.14 & 0.039 \\
		0.10 & 1.74 & 0.019 & 1.30 & 0.055 & 2.44 & 0.029 & 1.82 & 0.071 \\
		0.15& 2.50 & 0.030 & 1.52 & 0.037 & 2.80 & 0.039 & 2.21 & 0.079 \\
		0.20 & 2.77 & 0.036 & 1.96 & 0.049 & 3.09 & 0.042 & 2.67 & 0.10 \\
		0.25 & 2.99 & 0.040 & 2.60 & 0.096 & 3.37 & 0.045 & 3.33 & 0.14 \\
		0.30 & 3.21 & 0.046 & 2.76 & 0.080 & 3.62 & 0.067 & 3.56 & 0.43 \\
\hline
\end{tabular}}
\caption{Mean values of single voxel $R_1$ fits, $\overline{R_1}$, and standard deviation, $\sigma$ obtained via Inversion recovery (IR) and saturation recovery (SR) for solution and gel \MnCl phantoms.}
\label{table:R1-IR-SR-compare}
\end{table}

\begin{table}[!htb]\centering
\scalebox{0.75}{
\begin{tabular}{|c|cc|cc|cc|cc|}
\hline
& \multicolumn{2}{c|}{9-Point $\TR = \SI{4090}{ms}$} & \multicolumn{2}{c|}{12-Point $\TR = \SI{14000}{ms}$} & \multicolumn{2}{c|}{2-Point $\TR = \SI{4090}{ms}$} & \multicolumn{2}{c|}{2-Point $\TR = \SI{14000}{ms}$}\\
\hline
\textbf{\MnCl{} Conc} & \textbf{$\overline{\R1}$} & \textbf{$\sigma$} & \textbf{$\overline{\R1}$} & \textbf{$\sigma$}&\textbf{$\overline{\R1}$} & \textbf{$\sigma$} &\textbf{$\overline{\R1}$} & \textbf{$\sigma$}\\
(\si{mM}) & (\si{\per\second}) & (\si{\per\second}) & (\si{\per\second}) & (\si{\per\second})&(\si{\per\second})&(\si{\per\second})&(\si{\per\second})&(\si{\per\second})\\			
\hline
		0.00 & 1.00 & 0.044 & 0.79 & 0.098 & 1.03 & 0.044 & 1.078 & 0.045 \\
		0.05 & 1.46 & 0.12 & 1.27 & 0.15 & 1.48 & 0.12 & 1.59 & 0.16 \\
		0.10 & 1.69 & 0.11 & 1.53 & 0.15 & 1.73 & 0.12 & 1.82 & 0.11 \\
		0.15 & 2.65 & 0.20 & 2.55 & 0.21 & 2.67 & 0.20 & 2.75 & 0.21 \\
		0.20 & 3.08 & 0.19 & 3.21 & 0.23 & 3.09 & 0.20 & 3.24 & 0.27 \\
		0.25 & 3.24 & 0.18 & 3.30 & 0.19 & 3.28 & 0.20 & 3.38 & 0.22 \\
		0.30 & 4.38 & 0.37 & 4.50 & 0.47 & 4.42 & 0.38 & 4.54 & 0.42 \\
\hline
\end{tabular}}
\caption{Mean values of single voxel $R_1$ fits, $\overline{R_1}$, and standard deviation, $\sigma$ obtained for VFA methods for gel phantoms.  For the two-point method flip angles of $2^\circ$ and $15^\circ$ were used.  For the 9-point method with \TR=\SI{4.09}{ms} the flip angles were $2$, $3$, $4$, $6$, $8$, $10$, $12$, $13$ and $15$ degrees.  For the 12-point method with \TR=\SI{14}{ms} the flip angles were $2$, $5$, $10$, $15$, $20$, $30$, $50$, $60$, $70$, $80$ and $90$ degrees. }
\label{table:R1-VFA-compare}
\end{table}

\begin{table} \center
\scalebox{0.75}{\begin{tabular}{|c||*{4}{c|}} \hline
\MnCl{} Conc (mM)      & CI IR-SR Sol      & CI IR-SR Gel & CI IR-VFA2 Gel & CI SR-VFA2 Gel\\\hline
0.00 & $(0.62,0.63)$ & $(1.26, 1.27)$ & $(0.56,0.56)$ & $(-0.068,-0.062)$ \\
0.05 & $(0.53,0.56)$ & $(0.97, 0.97)$ & $(0.20,0.23)$ & $(-0.34,-0.32)$\\
0.10 & $(0.61,0.63)$ & $(0.44, 0.45)$ & $(0.75,0.76)$ & $(0.13,0.15)$ \\
0.15 & $(0.58,0.60)$ & $(0.98, 0.99)$ & $(0.15,0.17)$ & $(-0.44,-0.42)$\\
0.20 & $(0.40,0.42)$ & $(0.80, 0.81)$ & $(0.010,0.035)$ & $(-0.40,-0.37)$\\
0.25 & $(0.018,0.045)$ & $(0.37, 0.39)$ & $(0.16,0.18)$ & $(0.12,0.15)$\\
0.30 & $(0.024,0.11)$ & $(0.45, 0.47)$ & $(-0.77,-0.73)$ & $(-0.86,-0.77)$\\
\hline
\end{tabular}}
\caption{Pairwise t-test results (allowing for differences in variance) for \R1 distributions obtained with IR, SR and VFA 2-15 protocols for different phantoms show significant difference.  CI denotes the confidence intervals for the estimated difference in the means of the two \R1 distributions.  The $p$-values for all tests were $<0.001$, strongly rejecting the null hypothesis that the \R1 values come from the same distribution.}
\label{table:CI-R1-ttest}
\end{table}

\begin{figure}[!htb]\centering
\includegraphics[width=0.7\linewidth]{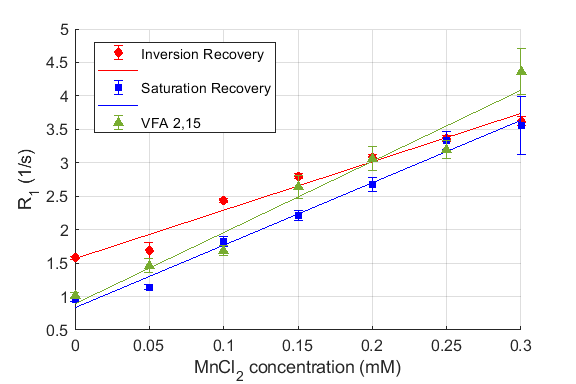}
\caption{Relaxivity plot \MnCl gel phantoms comparing IR, SR and 2-15 VFA protocols.}
\label{fig:Relaxivity1}
\end{figure}

Non-negligible differences in the \R1 values obtained for the same phantom are also evident in the relaxivity plots in Figure~\ref{fig:Relaxivity1}, which show that the \R1 values obtained for the IR protocol are consistently higher than the corresponding values obtained with the SR protocol, although the differences decrease slightly with increasing contrast agent concentration. This is the case for both the liquid and gel phantoms although the differences for the gel phantom are somewhat smaller.  Similarly, the \R1 values for the two-point VFA method start close to the values for the SR protocol for low concentrations of \MnCl but for high concentrations exceed the \R1 values obtained for the IR method.  These differences affect the slope of the linear regression line, which corresponds to the relaxivity of the contrast agent, resulting in different estimates for the relaxivity (see Table~\ref{table:relaxivity}) although the overlap of the 95\% confidence intervals of the estimates suggests that data for more phantoms with possibly a wider range of concentrations would be required to ascertain if the differences in relaxivity estimates are significant at the 95\% confidence level.

\begin{table}[!htb]\centering
\scalebox{0.8}{
\begin{tabular}{|c|c|c|c|}
\hline
Protocol &  relaxivity (95\% CI) & $y$-intercept (95\% CI) & $R^2$ of fit\\
         & \si{\per\second}  $\si{mM}^{-1}$  & \si{\per\second} & \\ 
\hline
Solution-IR &  6.16 (4.53,  7.78) & 1.41 (1.12, 1.71) & 0.95\\ 
Solution-SR &  8.43 (7.34,  9.54) & 0.31 (0.11, 0.51) & 0.99\\ 
Gel-IR      &  7.16 (5.64,  8.69) & 1.57 (1.29, 1.84) & 0.97\\ 
Gel-SR      &  9.22 (8.02, 10.43) & 0.84 (0.62, 1.06) & 0.99\\ 
9-Point     & 10.65 (8.29, 13.00) & 0.89 (0.46, 1.32) & 0.96\\ 
12-Point    & 11.91 (9.40, 14.45) & 0.64 (0.18, 1.10) & 0.97\\ 
2-Point \TR=4090  & 10.69 (8.41, 12.98) & 0.91 (0.49, 1.32) & 0.97\\ 
2-Point \TR=14000 & 10.87 (8.55, 13.18) & 0.98 (0.56, 1.40) & 0.97\\ 
 \hline
\end{tabular}}
\caption{Relaxivity fit coefficients for each phantom scan including 95\% confidence intervals, and $R^2$ fit error}
\label{table:relaxivity}
\end{table}

Further details can be found in Supplementary Material.  Figures~\ref{Fig:SolImages} and \ref{Fig:GelImages} show the IR and SR image series with ROIs selected and the \R1 map obtained for the liquid and gel phantoms, respectively.  Fig.~\ref{Fig:GelImages_VFA} shows the images for an ultrafast flip angle series with $\TR=\SI{4.09}{ms}$ and the corresponding \R1 maps obtained using two and nine flip angles, respectively.  For the IR and SR protocols, the corresponding transverse magnetisation signal (mean and standard deviation) as a function of \TR and \TI, respectively, with the best-fit curves, is shown in Figures~\ref{Fig:SolPlots} and \ref{Fig:GelPlots} for different concentrations of the contrast agent, for both liquid and gel contrast phantoms.  Although only the mean signal fits are shown, they are representative of the single voxel fits.  The quality of the non-linear fits of the saturation and inversion recovery curves obtained was high: the majority of the single voxel fits have $R^2$ values $\ge 0.99$, and the percentages of single voxel fits with $R^2<0.95$ are low at $1.2\%$, $2.5\%$, $15.1\%$ $0.02\%$ for IR gel fits, SR gel fits, IR solution fits and SR solution fits, respectively.

Although differences in signal intensity are visible in the images, it is difficult to reliably distinguish all contrast phantoms based on a single image.  The \R1 maps differentiate the contrast phantoms much better.  For all methods tested, each phantom is mutually distinguishable, in that the pairwise t-test for the \R1 distributions for any two phantoms rejects the null hypothesis that they come from the same distribution at the $p=0.01$ level, even when allowing for unequal variances of the distributions.  

\begin{figure}[!htb]\centering
\subfloat[IR Series (\MnCl Solutions)]{\includegraphics[width=0.49\linewidth]{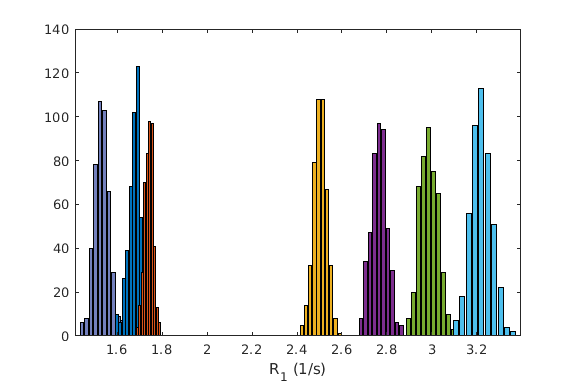}}
\subfloat[SR Series (\MnCl Solutions)]{\includegraphics[width=0.49\linewidth]{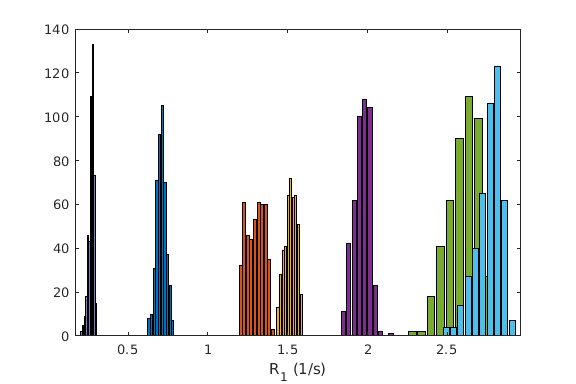}}\\
\subfloat[IR Series (\MnCl Gels)]{\includegraphics[width=0.49\linewidth]{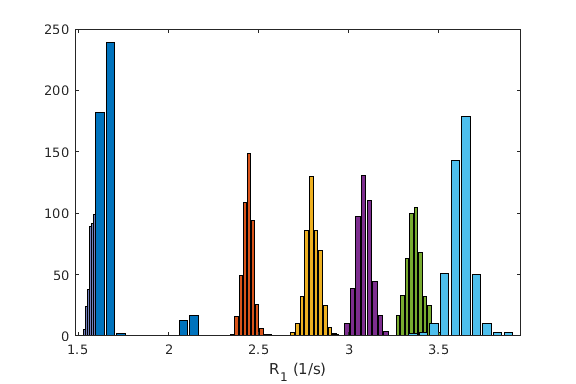}}
\subfloat[VFA Series (\MnCl Gels)]{\includegraphics[width=0.49\linewidth]{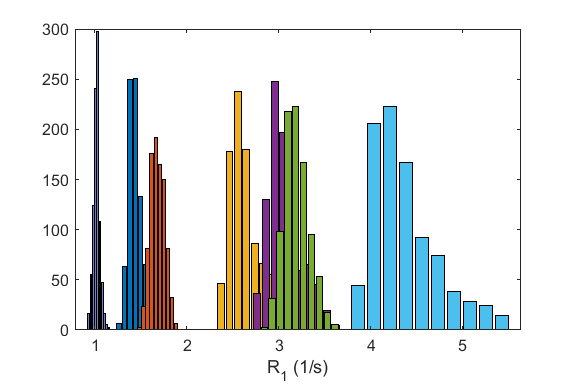}}
\caption{Histogram of \R1 values for obtained for \MnCl phantoms with various methods. Each phantom is plotted separately; \SI{0.00}{mM} (light-purple), \SI{0.05}{mM}(dark-blue), \SI{0.10}{mM}(orange), \SI{0.15}{mM}(yellow), \SI{0.20}{mM}(dark-purple), \SI{0.25}{mM}(green), \SI{0.30}{mM}(light-blue).}
\label{fig:histogram}
\end{figure}

Analysis of the distribution of \R1 values for different phantoms and methods in Fig.~\ref{fig:histogram}, however, shows that there are variations in the shape of the distributions for individual phantoms and the degree of overlap for different phantoms.  Considering the overlap between the distributions, the SR protocol appears slightly worse in discriminating the phantoms with the highest concentration of \MnCl while our IR protocol appears slightly worse for the lower contrast phantoms.  Comparing the distributions for the IR and two-point VFA method also shows that the distributions for the latter are much broader with greater overlap of the distributions for different phantoms. This is also reflected in the larger standard deviation of the \R1 values obtained with VFA methods (see Tables \ref{table:R1-IR-SR-compare}, \ref{table:R1-VFA-compare}).

\subsection*{In vivo: \R1 Mapping for Brains}

\begin{figure}[!htb]\centering
\includegraphics[width=1.1\textwidth]{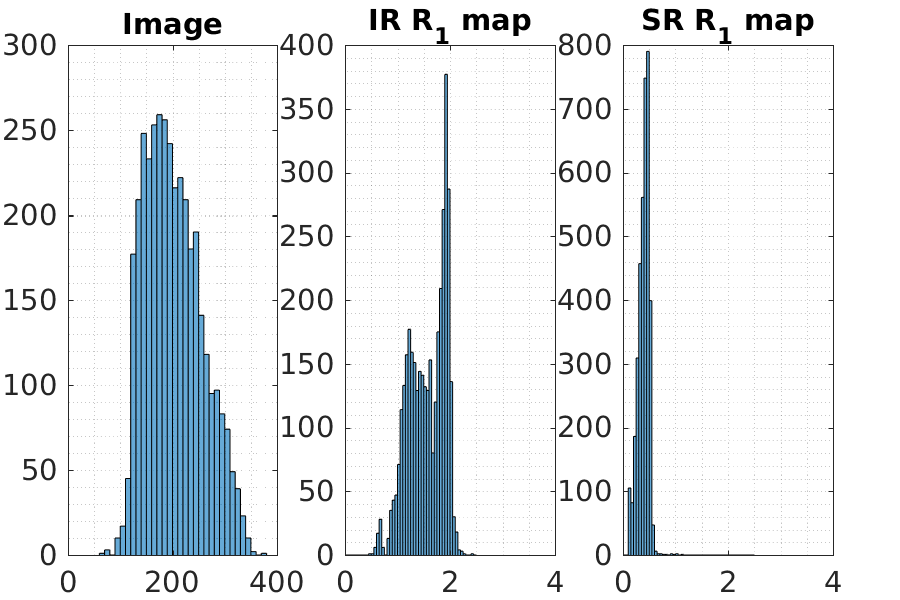}
\caption{Brain histograms for spin echo signal, \R1 from IR, and \R1 from SR. Notice IR is the only one to present a bimodal distribution. This allows us to separate the grey and white matter regions using thresholidng and connected component analysis.}
\end{figure}

\begin{table}[!htb]
\centering
\scalebox{0.8}{
\begin{tabular}{|c|cc|cc|cc|}
\hline
& \multicolumn{2}{c|}{Saturation recovery} &	\multicolumn{2}{c|}{Inversion recovery}&& \\			
\hline
ROI & $\overline{R_1}$ (\si{mM}) & $\sigma$ (\si{mM}) & $\overline{R_1}$ (\si{mM}) & $\sigma$ (\si{mM}) & t-test p-value & t-test CI\\
\hline
    Cerebellum & 0.54 & 0.05 & 0.65 & 0.057 & 2.39e-13 & (0.089, 0.14) \\
	& 0.55 & 0.068 & 0.75 & 0.079 & 1.78e-17 & (0.16, 0.23) \\
    & 0.58 & 0.067 & 0.72 & 0.063 & 4.11e-14 & (0.12, 0.18) \\
     & 0.53 & 0.090 & 0.69 & 0.089 & 4.98e-10 & (0.11, 0.19) \\
\hline	
     Thalamus& 0.46 & 0.061 & 0.79 & 0.11  & 9.74e-22 & (0.30, 0.38) \\
	 & 0.58 & 0.035 & 0.92 & 0.092 & 2.18e-24 & (0.30, 0.37) \\
	 & 0.65 & 0.074 & 0.90 & 0.059 & 2.09e-24 & (0.22, 0.28) \\
	 & 0.64 & 0.057 & 0.98 & 0.089  & 2.90e-27 & (0.30, 0.38) \\
\hline\hline
    Grey matter& 0.31 & 0.088 & 1.24 & 0.21 & 0.00 & (0.94, 0.92) \\
	& 0.33 & 0.11 & 1.21 & 0.27 & 0.00 & (0.89, 0.87) \\
	& 0.42 & 0.17 & 1.23 & 0.24 & 0.00 & (0.82, 0.80) \\
	& 0.41 & 0.27 & 1.21 & 0.28 & 0.00 & (0.82, 0.79) \\
\hline
    White matter & 0.47 & 0.048 & 1.90 & 0.091 & 0.00 & (1.43, 1.42) \\
	& 0.48 & 0.064 & 1.89 & 0.091 & 0.00 & (1.41, 1.40) \\
	& 0.63 & 0.11  & 1.89 & 0.090 & 0.00 & (1.26, 1.25) \\
	& 0.60 & 0.066  & 1.82 & 0.066  & 0.00 & (1.23, 1.22) \\
\hline
\end{tabular}
}
\caption{Mean values of single voxel $R_1$ fits, $\overline{R_1}$, and standard deviation, $\sigma$ obtained via Inversion recovery(IR) and saturation recovery(SR) for volunteer cohort. The results are separated by region, then ordered by volunteer; volunteer 1, volunteer 2 scan 1, volunteer 2 scan 2, volunteer 3.}
\label{table:brain}
\end{table}

Table~\ref{table:brain} shows a summary of the \R1 mapping results for different brain regions and four in vivo data sets, corresponding to three different volunteers with volunteer 2 being scanned twice on two consecutive days. IR sequences consistently report a higher \R1 (lower \T1), irrespective of region or volunteer. Table~\ref{table:brain} also shows the t-test $p$-values, an measure of the probability of the observed separation being a result of a single distribution. All reported p-values for the in-vivo SR and IR comparisons fell below \SI{4.977e-10}{}, indicating a strong separation of \R1 distributions.  IR scans test-retest stability, with  $\overline{\R1}$ for the repeat patient reproduced to at least one decimal place in all cases.  However SR protocols do not exhibit the same stability, with the inter-participant variation on the same order as the intra-participant variation. Contrast is maintained in both sequences, but absolute quantification of \R1 appears more stable for the ``gold standard'' IR sequence.

\section*{Discussion}\label{S:discussion}

Although each method of \R1 mapping is able to discriminate between the different contrast phantoms, the spread and overlap of the distributions obtained for a single phantom varies depending on the method used.  Large differences in the means and distributions of the \R1 values for same phantom or tissue ROI, obtained with different methods, suggest that \R1 maps acquired with different methods cannot be reliably compared.  Indeed, the variation of \R1 values for a particular ROI obtained with different acquisition methods is larger than the regional variation, which suggests that differences in the \R1 quantification method may be a major factor in the large variability of the values reported in the literature, such as in the review by Bojorquez \etal, which reported \R1 of \SIrange{0.55}{1.03}{\per\second} for grey matter and \SIrange{0.70}{1.33}{\per\second} for white matter.  The \textit{in vivo} \R1 values for grey matter obtained from IR in our volunteer group lie within this range but the SR values are lower. For white matter the \R1 values for our volunteer group were found to lie above this range for IR but below for SR.

Large disparities between the IR and SR \R1 values are a consistent finding of this study.  For all regions and volunteers, the differences between SR and IR values fell outside the bounds of standard deviation, with t-test $p$-values ranging from $0$ to $10^{-10}$, suggesting strongly statistically significant differences.  IR \R1 values are found to be consistently higher than SR values for the same regions, in agreement with the findings of the phantom study.  SR \R1 values also showed poor reproducibility in grey and white matter quantification, with inter-participant variation on the same order as intra-participant variation.  Conversely, the IR method gives consistent intra-participant results, while maintaining inter-participant variation.

Our findings further indicate that the choice of acquisition method affects not only \R1 values, but also contrast agent relaxivity, $r_1$.  The $r_1$ value of \SI{6.397}{Hz/mM} for \MnCl solutions reported in~\cite{Thangavel2017} falls within the 95\% confidence interval of the $r_1$ obtained with our IR protocol.  Specifically, our $r_1^{\rm IR}$ is about 4\% lower than the value reported in~\cite{Thangavel2017}, while  $r_1^{\rm SR}$ is almost 32\% higher, and the discrepancies are even greater for the VFA method.  \MnCl{} relaxivity for the gel phantoms was found to vary between between $7.16 (5.64, 8.69)$ and $11.91 (9.40, 14.45)$ \si{Hz/mM} for IR and the 12-point VFA respectively.  While gel phantoms are expected to have higher relaxivity than solutions, $r_1^{\rm IR}$ of the 1\% agar gels is approximately 16\% higher than the value for the solutions, while the two-point VFA method (2-15) gives a value approximately 67\% higher than the value reported in~\cite{Thangavel2017}.  The similarity of the IR relaxivity in our study and the value reported by Thangeval \etal could be due to the fact that the latter study used an IR protocol on a 3 T Siemens scanner similar to ours, and we have found excellent consistency in our phantom data when the same protocol is repeated with the same phantoms on the same hardware. 

Discrepancies in both \R1 and $r_1$ values obtained using different protocols have implications for many clinical applications.  For instance, in DCE MRI, $R_1$ values obtained using a two-point (typically 2-15) VFA method are often combined with known relaxivities of the contrast agent to model contrast agent uptake.  Given known relaxivities of free Gadolinium in tissue such as skin, bone or brain \R1 mapping could potentially also to be used for absolute quantification of Gadolinium in such tissues.  Our results suggest that ideally \R1 mapping should be done with an IR protocol.  However, in practice, accuracy and precision must be balanced with the efficiency of the acquisition protocol and VFA methods are often the only feasible option for DCE MRI where rapid acquisition is essential.  In this case, the same protocol should be used to establish the $r_1$ of the contrast agent to reduce over- or underestimation of contrast agent uptake or Gadolinium deposits. 

The significant differences in both longitudinal relaxation rates and relaxivity obtained suggest that the underlying simple relaxation models do not accurately describe the dynamics and relaxation characteristics of the systems studied when subjected to different RF pulse sequences. More complex relaxation models involving multiple compartments such as a solid and free water pool relaxing at different rates have been proposed.  Specifically, it has been argued that in SR experiments both the free water and solid pool are saturated, while in IR experiments the free water pool is inverted while the solid pool is saturated.  Exchange between the solid and free water pool has been shown to lead to different recovery curves~\cite{Malik2018}.  This further exacerbates the methodological dependence observed in the literature~\cite{Novikov2018}.  However, while the existence of solid pools and magnetisation transfer effects may explain some of the differences in the \textit{in vivo} $R_1$ results, such effects are not applicable to contrast agent solutions and expected to be small for 1\% agar gels~\cite{Henkelman1993}.  Thus, this model cannot explain the large differences in both longitudinal relaxation rates and relaxivities obtained for the phantom data.  Furthermore, $R^2$ values close to $1$, especially for the phantom data, suggest that the mono-exponential model used to fit the relaxation curves is a good fit for the data.  This does not rule out that the pulse sequences applied can alter the effective $R_1$ by other mechanisms.

This raises the question what is really being measured in different experiments, whether there is a true \R1 value, and which experimental and analytical methods are best suited to elucidate the underlying dynamics.  While it could be argued that it does not matter if we measure the true \R1, or even if there is a true \R1, provided that we apply a protocol that is effective at discriminating different tissue types and gives results that are consistent and reproducible with a small margin of error, this is problematic for quantitative MRI and quantification of biomarkers, as the characteristics of a true biomarker should not be dependent on the MRI protocol or hardware characteristics, aside from physical variables such as field strength.

This study was necessarily limited in scope and future work is required to address issues such as the effect of the various combinations of flip angles for VFA methods, other sequences such as MPRAGE, MP2RAGE, Look-Locker etc.  Acquisition of more phantom and \textit{in vivo} data to corroborate the results and perhaps more detailed modelling of the RF pulses sequences applied are also desirable.

\FloatBarrier
\section*{Conclusion}
\label{S:conclusion}

While the temporal consistency of phantom data and \textit{in-vivo} IR maps is promising, the marked separation of IR and SR maps and large differences in both $R_1$ and contrast agent relaxivity $r_1$ values obtained using different protocols is troubling, and emphasises the need for standardisation of protocols where possible. Comparisons of \R1 measurements across acquisition and analysis protocols should be done tentatively, if at all, to avoid invalid conclusions.

\section*{Acknowledgements}
We thank Rhodri Evans, Jonathan Phillips and Kenith Meissner for helpful discussions and suggestions.   

\FloatBarrier

\clearpage  
\setcounter{page}{1}
\setcounter{figure}{0}
\setcounter{table}{0}
\setcounter{section}{0}
\renewcommand{\thepage}{S\arabic{page}}
\renewcommand{\thesection}{S\arabic{section}}
\renewcommand{\thetable}{S\arabic{table}}
\renewcommand{\thefigure}{S\arabic{figure}}
\renewcommand{\figurename}{Supplemental Material, Figure}

\section*{Supplementary Material}\label{appendix:R1Phantom}

\begin{figure*}[!h]
	\subfloat[IR Series]{\includegraphics[width=0.43\textwidth]{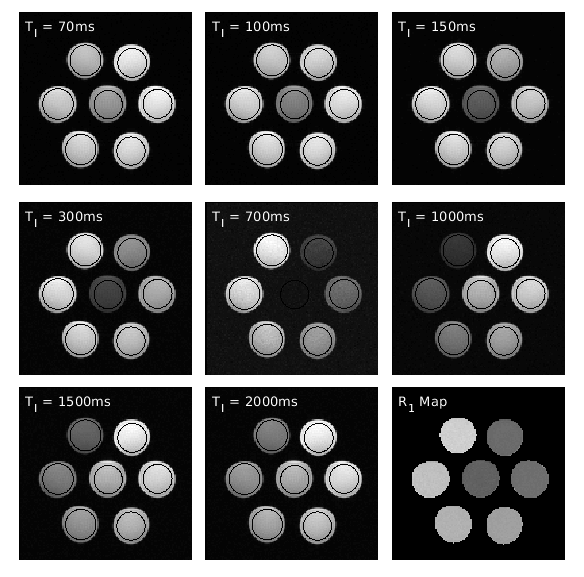}}
	\subfloat[SR Series]{\includegraphics[width=0.57\textwidth]{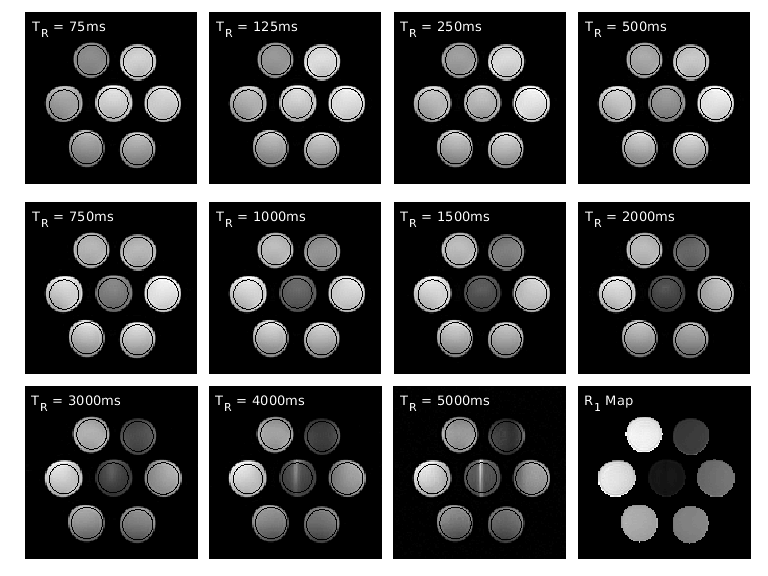}}
	\caption{Inversion recovery and saturation recovery images for \MnCl solutions.} \label{Fig:SolImages}
	\subfloat[IR Graphs]{\includegraphics[width=0.5\textwidth]{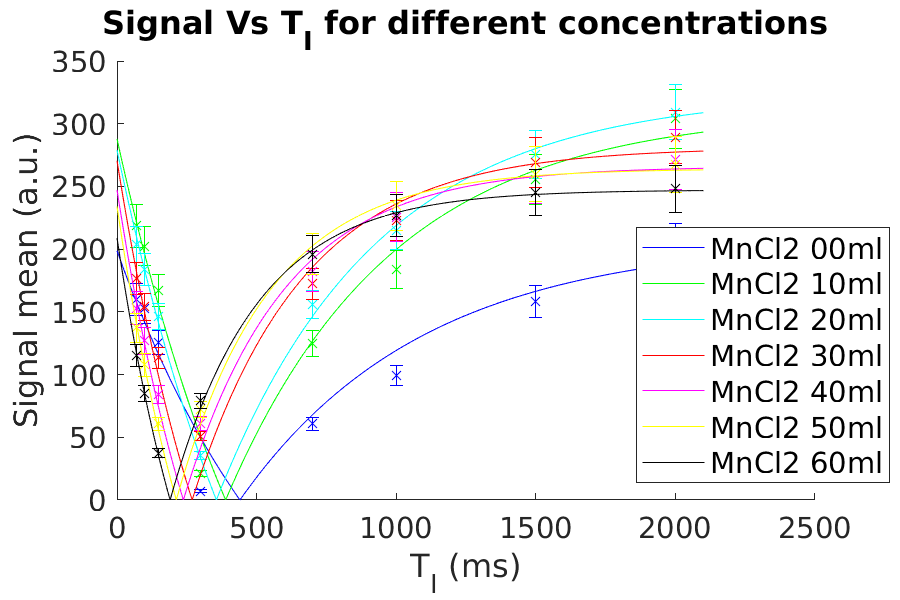}}
	\subfloat[SR Graphs]{\includegraphics[width=0.5\textwidth]{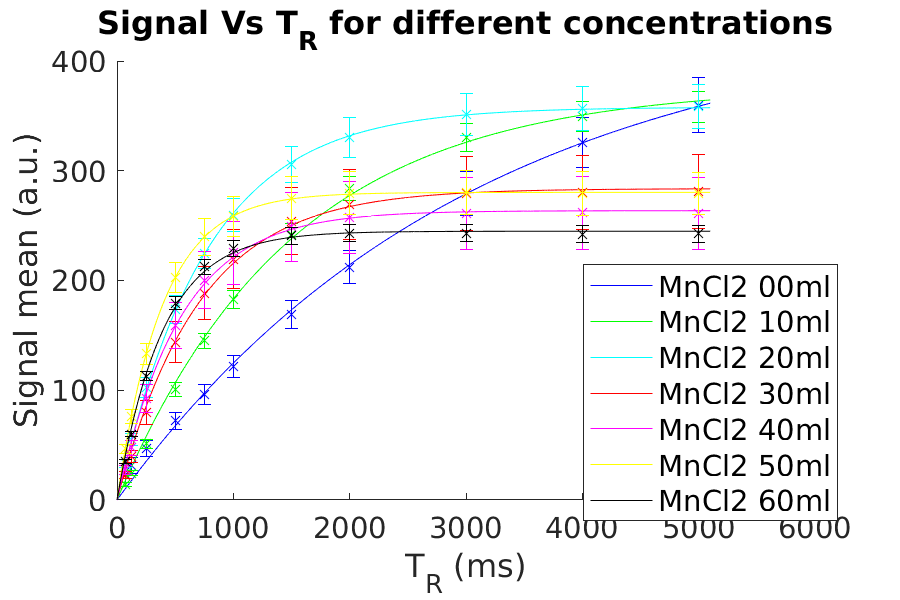}}
	\caption{Mean signal vs $T_I$ ($T_R$) graphs for \MnCl contrast solutions.  The error bars indicate the variation of the signal over the ROI.  The solid lines are the curves of best fit.} \label{Fig:SolPlots}
\end{figure*}

\begin{figure*}[!h]
	\subfloat[IR Series]{\includegraphics[width=0.43\textwidth]{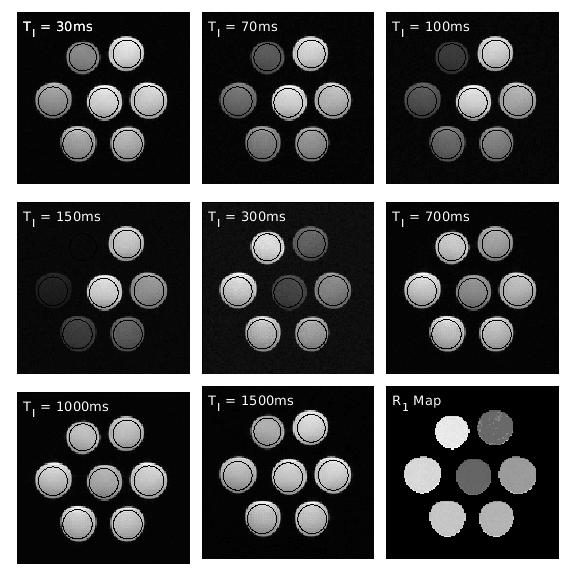}}
	\subfloat[SR Series]{\includegraphics[width=0.57\textwidth]{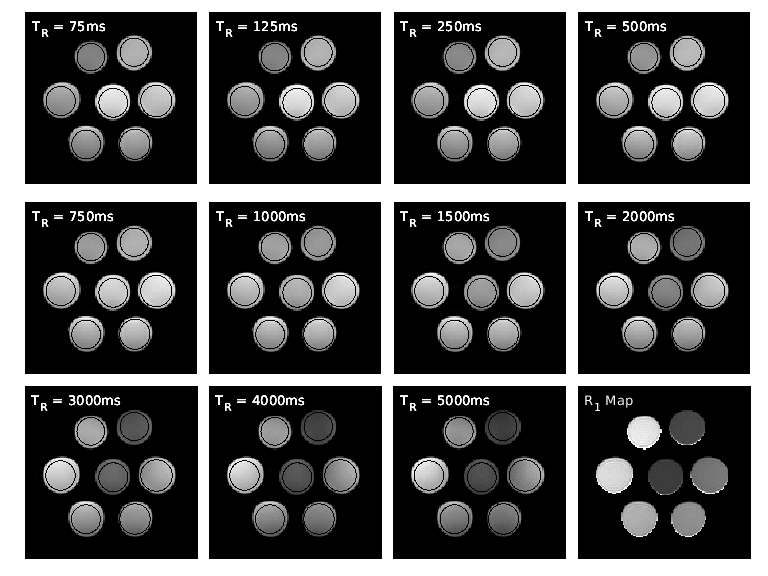}}
	\caption{Inversion recovery and saturation recovery images for \MnCl gel phantoms.} \label{Fig:GelImages}
	\subfloat[IR Graphs]{\includegraphics[width=0.5\textwidth]{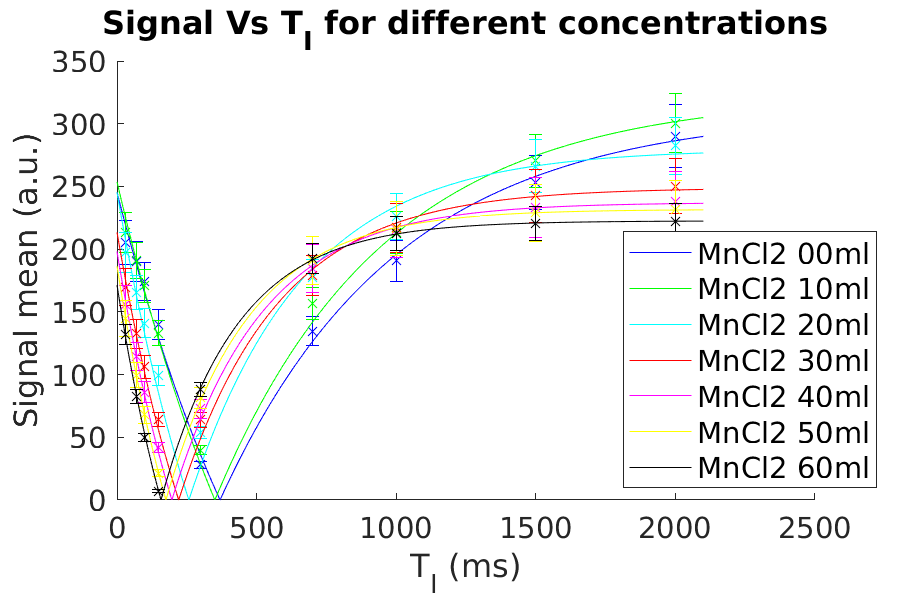}}
	\subfloat[SR Graphs]{\includegraphics[width=0.5\textwidth]{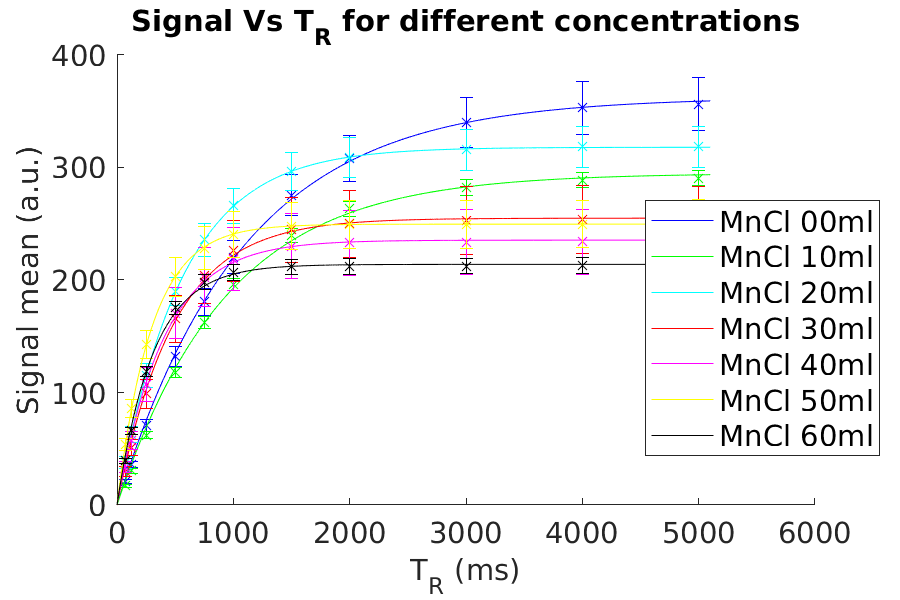}}
	\caption{Mean signal vs $T_I$ ($T_R$) graphs for \MnCl contrast gel phantoms.  The error bars indicate the variation of the signal over the ROI.  The solid lines are the curves of best fit.}
	\label{Fig:GelPlots}
\end{figure*}

\begin{figure*}[!h]\centering
	\includegraphics[width=0.7\textwidth]{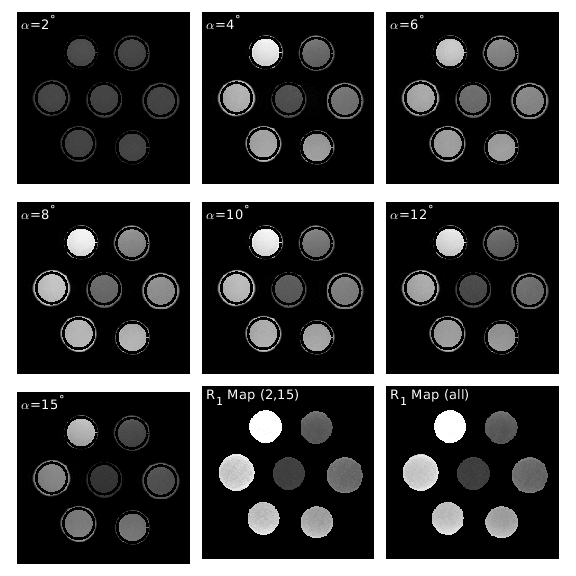}
	\caption{VRA series images with \R1 maps obtained using only two flip angles ($2$ and $15$ degrees) and 9 flip angles ($2,3,4,6,8,10,12,13,15$ degrees) for \MnCl gels.} 
	\label{Fig:GelImages_VFA}
\end{figure*}

\begin{figure}[!htb]\centering
	\includegraphics[width=1.1\textwidth]{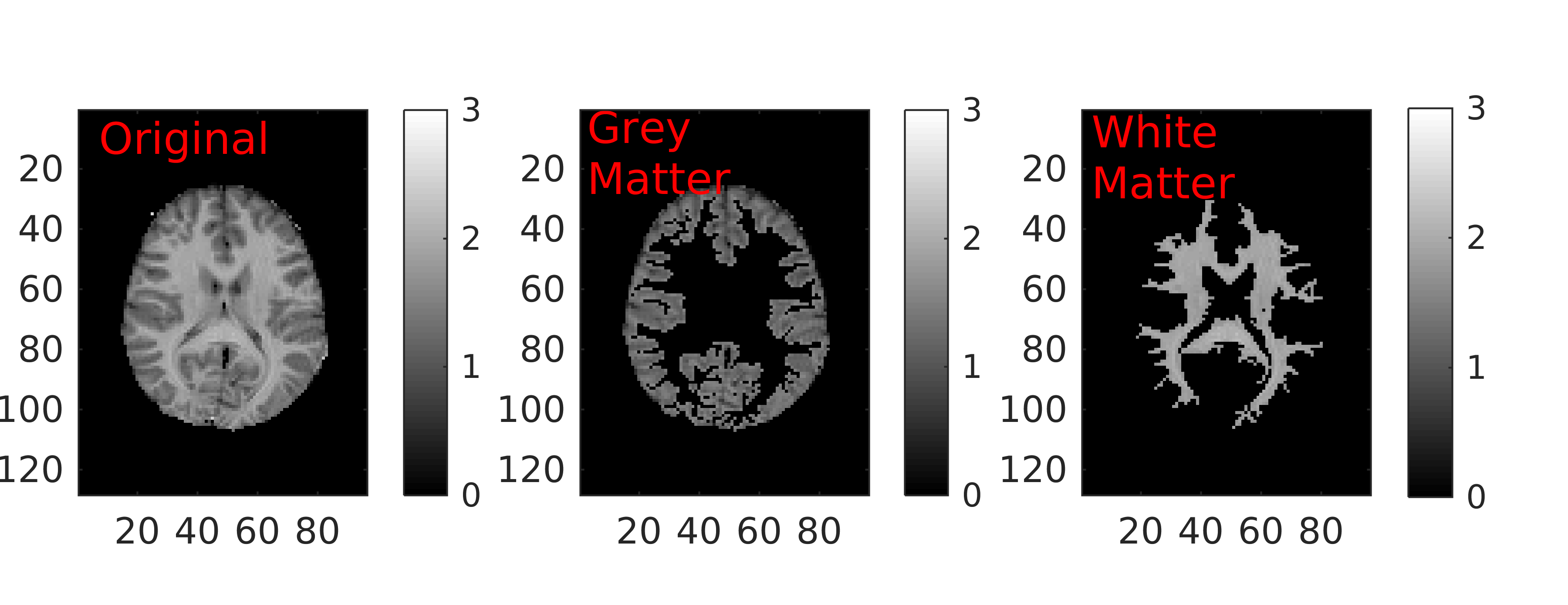}
	\caption{Example of grey and white matter delineation for one volunteer. The thalamus region is excluded, and the grey and white matter separated using image processing.}
	\label{fig:GMWMDelin}
\end{figure}

\FloatBarrier

\end{document}